\documentclass{iopart}
\usepackage[colorlinks=true,citecolor=blue]{hyperref}
\usepackage{graphicx}
\usepackage{iopams}

\begin{document}
%\begin{CJK*}{UTF8}{}             % use CJK unicode characters
%\preprint{INT PUB 08-22}

\title[Sources for elements heavier than Fe]{Diverse, 
massive-star-associated sources for elements heavier than Fe 
and the roles of neutrinos}

\author{Yong-Zhong Qian$^{1,2}$}
\address{$^1$ School of Physics and Astronomy, %
University of Minnesota, Minneapolis, MN 55455, USA}
\address{$^2$ Center for Nuclear Astrophysics, %
Shanghai Jiao Tong University, Shanghai 200240, China}
\ead{\mailto{qian@physics.umn.edu}}

\begin{abstract}
Massive-star-associated models for production of elements heavier than Fe
are reviewed. The important roles of neutrinos in many of these models are 
discussed along with uncertainties in the relevant neutrino physics. 
Data on elemental abundances in metal-poor stars are presented and their
constraints on diverse sources for elements heavier than Fe are emphasized. 
\end{abstract}
 
\maketitle

\section{Introduction}
The inventory of those elements heavier than Fe in the solar system mainly came
from two types of neutron-capture processes: the $s$ (slow) process where the
neutron-capture rates of the produced unstable nuclei are slower than the
$\beta$-decay rates of the same nuclei and the $r$ (rapid) process where the opposite 
is true \cite{1957RvMP...29..547B,1957PASP...69..201C}. In addition to the crucial
input from nuclear physics, understanding these two processes
requires detailed knowledge of the conditions in the relevant astrophysical 
environments. Neutron-capture rates are directly proportional to 
the neutron (number) density $n_n$. These rates also depend on the 
temperature $T$ through the energy dependence of the cross sections
(especially due to resonances) and through the relative velocity between neutrons
and nuclei. The presence of nuclear excited states introduces
dependence on $T$ for both neutron-capture (e.g., \cite{2010PhRvC..82a5804F})
and $\beta$-decay rates (e.g., \cite{1980ApJS...42..447F}). In some environments,
$\beta$-decay rates are affected by the ionization states of host atoms 
(e.g., \cite{1987ADNDT..36..375T}) or by Pauli blocking of the emitted electrons
(e.g., \cite{1980ApJS...42..447F}). Most important,
how neutrons and the seed nuclei capturing them are provided for
a neutron-capture process and more generally, how the overall nucleosynthesis 
proceeds as a function of time are determined by the detailed, often dynamic 
conditions in the relevant environments. Finally, in order to compare the contributions 
from different environments sited in various astrophysical events, we also need to know 
the total amount of neutron-capture elements produced in an event and the associated 
event rate over the Galactic history.

All isotopes of Tc are radioactive with $^{98}$Tc having the longest half-life of
$4.2\times 10^6$~yr. The discovery of Tc in the spectra of stars now known to be 
in the asymptotic-giant-branch (AGB) stage of their evolution
\cite{1952ApJ...116...21M} firmly established that the main
$s$ process occurs in AGB stars of $\sim 1$--$3\,M_\odot$ ($M_\odot$ being
the mass of the sun; e.g., \cite{2011RvMP...83..157K}).
The search for the astrophysical sites of the $r$ process so far has been 
inconclusive. Nevertheless, it has
benefited greatly from models of massive ($\gtrsim 8\,M_\odot$) star evolution
(e.g., \cite{2002RvMP...74.1015W}), 
core-collapse supernovae (e.g., \cite{2012ARNPS..62..407J}),
and neutron star mergers (e.g., \cite{1999ApJ...525L.121F,2011ApJ...738L..32G}), 
from recognition of the importance of
neutrinos to these models (e.g., \cite{1985ApJ...295...14B,2003MNRAS.342..673R}), 
and from observations of elemental abundances in stars
formed over the Galactic history (e.g., \cite{2008ARA&A..46..241S})
as well as meteoritic measurements of radioactive isotopes present in the early 
solar system (e.g., \cite{1996ApJ...466L.109W,2006NuPhA.777....5W}).
The above efforts not only have improved our understanding of the $r$ process,
but also have led to discoveries of new processes that make elements heavier than 
Fe not in the same way as the classical $r$ or $s$ process
(e.g., \cite{1992ApJ...395..202W,2006PhRvL..96n2502F}).

Recent progress in the understanding of the $s$ process has been reviewed by
Ref.~\cite{2011RvMP...83..157K}. This paper focuses on models of the $r$ process 
and other related processes for producing elements heavier than Fe, all of which
are associated with massive stars. Emphases are on the diversity of the relevant
environments, the roles of neutrinos in quite a few of them, and the constraints from
observations of metal-poor stars formed in the early Galaxy. 
Extensive discussion of nuclear physics issues and of astrophysical models with
different emphases can be found in earlier reviews (e.g., 
\cite{1991PhR...208..267C,2003PrPNP..50..153Q,2007PhR...450...97A}).

\section{Astrophysical models of the $r$ process and related nucleosynthesis
\label{sec-model}}
By definition, an $r$ process requires 
\begin{equation}
n_n\langle v\sigma_{n,\gamma}(Z,A)\rangle\gtrsim\lambda_\beta(Z,A), 
\end{equation}
where $v$ is the relative velocity and $\sigma_{n,\gamma}(Z,A)$ is the corresponding 
cross section for neutron capture by the nucleus $(Z,A)$ with proton number $Z$ and 
mass number $A$, $\langle v\sigma_{n,\gamma}(Z,A)\rangle$ is the neutron-capture 
rate coefficient of this nucleus averaged over the relevant thermal distributions, and 
$\lambda_\beta(Z,A)$ is its $\beta$-decay rate. A crude estimate of the neutron density
required for an $r$ process is 
\begin{equation}
n_n\gtrsim 10^{18}\left[\frac{10^{-17}\ {\rm cm}^3\ {\rm s}^{-1}}
{\langle v\sigma_{n,\gamma}(Z,A)\rangle}\right]
\left[\frac{\lambda_\beta(Z,A)}{10\ {\rm s}^{-1}}\right]\ {\rm cm}^{-3},
\label{eq-nn}
\end{equation}
where nominal values of $\langle v\sigma_{n,\gamma}(Z,A)\rangle$ and
$\lambda_\beta(Z,A)$ have been indicated. 

For convenience of discussion, we distinguish
two generic types of $r$-process models: (1) a hot $r$-process occurring at
$T\gtrsim 10^9$~K, for which neutron emission by photo-disintegration is important, and
(2) a cold $r$ process occurring at $T\ll 10^9$~K, for which photo-disintegration can be 
ignored. A hot $r$ process tends to go through a phase with
$n_n\gtrsim 10^{20}$~cm$^{-3}$ and $T\gtrsim 10^9$~K, for which
a statistical $(n,\gamma)$-$(\gamma,n)$ equilibrium is achieved between
neutron capture and photo-disintegration.
In this case, the $r$-process path is determined by the
``waiting-point'' nuclei with neutron separation energies favored by the
$n_n$ and $T$ of the environment and is independent of the detailed 
dynamics of neutron capture and photo-disintegration (e.g.,
\cite{1991PhR...208..267C,2003PrPNP..50..153Q,2007PhR...450...97A}). 
Upon the $\beta$-decay of the waiting-point nucleus with proton number $Z$,
its daughter nucleus immediately captures several neutrons to become 
the waiting-point nucleus with proton number $Z+1$. Except for
a small number of nuclei with closed neutron shells,
waiting-point nuclei have typical $\beta$-decay rates of 
$\lambda_\beta\gg 10$~s$^{-1}$. Consequently,
neutrons are rapidly consumed in the $(n,\gamma)$-$(\gamma,n)$-equilibrium
phase of a hot $r$ process. In the absence of fission, the total number of nuclei 
(excluding neutrons) is conserved. A useful quantity to characterize a hot $r$ 
process is the number of neutrons per seed nucleus, or the neutron-to-seed ratio 
$n/s$, at the beginning of the process. When all the neutrons are captured, 
the abundance pattern produced by a hot $r$ process is required by mass 
conservation to have the average mass number
\begin{equation}
\langle A\rangle = \langle A_s\rangle + n/s,
\label{eq-n2s}
\end{equation}
where $\langle A_s\rangle$ is the average mass number of the seed nuclei.
For $\langle A_s\rangle + n/s\gtrsim 260$, the heaviest nuclei produced
are unstable to fission. This increases the total number of 
nuclei and Eq.~(\ref{eq-n2s}) no longer applies. The fission fragments become the
new seed nuclei to capture neutrons. In the limit where a cyclic flow is established
between the fission fragments and the fissioning nuclei (i.e., fission cycling occurs), 
a steady abundance pattern results independently of the initial seed nuclei and $n/s$.

As photo-disintegration plays no role in a cold $r$ process,
a useful quantity to characterize this process is the time integral of $n_n$, 
or the total neutron ``exposure''
\begin{equation}
{\cal{T}}_n=\int_{t_i}^{t_f}n_n(t)dt,
\end{equation}
where $t_i$ and $t_f$ correspond to the time at the beginning and the end 
of the process, respectively. In the absence of fission, the average mass number 
of the nuclei produced by a cold $r$ process can be estimated as
\begin{equation}
\langle A\rangle\sim\langle A_s\rangle +
{\cal{T}}_n\overline{\langle v\sigma_{n,\gamma}\rangle},
\label{eq-nexp}
\end{equation}
where $\overline{\langle v\sigma_{n,\gamma}\rangle}$ is some average of
the neutron-capture rate coefficients for the nuclei involved. Models
of a cold $r$ process typically have relatively low $n_n$ and the corresponding
${\cal{T}}_n$ may be insufficient to consume all the available neutrons. This
is why ${\cal{T}}_n$ is more relevant than $n/s$ for such a process. A successful
cold $r$ process occurs when moderately high $n_n$ can be sustained for
long enough to give ${\cal{T}}_n$ required for producing heavy $r$-process nuclei.
(In contrast, a hot $r$ process typically has so high $n_n$ that
all available neutrons are captured. A successful hot $r$ process occurs when
enough neutrons are present initially to give a large $n/s$ required for producing 
heavy $r$-process nuclei.) If a cold $r$ process has $\langle A_s\rangle+
{\cal{T}}_n\overline{\langle v\sigma_{n,\gamma}\rangle}\gtrsim 260$, then fission
must be taken into account and Eq.~(\ref{eq-nexp}) no longer applies. 
If fission cycling could occur, a steady abundance pattern would be produced
but it may differ from that in the case of a hot $r$ process as the
nuclei involved in the cyclic flow are different for the two cases.

While the dynamics of an $r$ process depends on the occurrences of 
all the relevant nuclear reactions under the conditions of a specific
environment, the above discussion gives an approximate description of
the outcome based on the neutron-to-seed ratio $n/s$ for a hot $r$ process 
and the total neutron exposure ${\cal{T}}_n$ for a cold one. In terms of
specific features, because those unstable nuclei with closed neutron shells 
at $N=50$, 82, and 126 on the $r$-process path are more stable than nearby 
nuclei, they are accumulated to produce the signature $r$-process abundance 
pattern with peaks at $A\sim 80$, 130, and 195.
In contrast, the $s$-process pattern has 
peaks at $A=88$, 138, and 208 corresponding to the stable nuclei with the
$N=50$, 82, and 126 closed neutron shells. Other processes producing
nuclei heavier than Fe do not have as clear a signature abundance pattern
as either the $r$ or $s$ process.
Below we will focus on how the neutron-to-seed ratio or the total neutron 
exposure is determined in different astrophysical models for the $r$ process.

\subsection{Expansion from nuclear statistical equilibrium\label{sec-nse}}
A generic model of a hot $r$ process is based on expansion of material
from an initial state of high $T$ and high (mass) density $\rho$, for which
all strong and electromagnetic reactions among free nucleons and nuclei
occur so fast that the relative abundances of nuclei are determined by 
nuclear statistical equilibrium (NSE) independently of the detailed 
dynamics of these reactions. This initial composition and the
subsequent evolution of $T$ and $\rho$ determine the neutron-to-seed
ratio and ultimately the abundance pattern produced by the $r$ process.

In NSE, the number of nuclei $(Z,A)$ 
per baryon, or the number fraction $Y(Z,A)$ is given by
\begin{eqnarray}
Y(Z,A)&=&\frac{G(Z,A)A^{3/2}}{2^A}Y_n^{A-Z}Y_p^Z\nonumber\\
&&\times\left(\frac{2\pi\hbar^2}{m_ukT}\right)^{3(A-1)/2}
\left(\frac{\rho}{m_u}\right)^{A-1}\exp\left[\frac{B(Z,A)}{kT}\right],
\label{eq-nse}
\end{eqnarray}
where $G(Z,A)$ is the partition function of the corresponding nucleus, 
$B(Z,A)$ is its binding energy, $Y_n$ and $Y_p$
are the number fraction of neutrons and protons, respectively, $\hbar$ is the Planck
constant, $m_u$ is the atomic mass unit, and $k$ is the Boltzmann constant.
The composition in NSE is also subject to constraints from conservation of
baryon number and electric charge:
\begin{eqnarray}
Y_n+Y_p+\sum_{(Z,A)}AY(Z,A)&=&1,\label{eq-bnum}\\
Y_p+\sum_{(Z,A)}ZY(Z,A)&=&Y_e,\label{eq-ye}
\end{eqnarray}
where the sums extend over all nuclei (excluding free nucleons) and $Y_e$
is the net number of electrons per baryon or the electron fraction.
Given all the nuclear input for determining $G(Z,A)$ and $B(Z,A)$,
the NSE composition for specific values of $T$, $\rho$, and $Y_e$ can be 
calculated from Eqs.~(\ref{eq-nse}), (\ref{eq-bnum}), and (\ref{eq-ye}).

It is instructive to rewrite Eq.~(\ref{eq-nse}) as
\begin{eqnarray}
Y(Z,A)&=&\frac{G(Z,A)A^{3/2}}{2}Y_n^{A-Z}Y_p^Z[f(T,\rho;Z,A)]^{A-1},
\label{eq-nse1}\\
f(T,\rho;Z,A)&\equiv&\left(\frac{2\pi^2}{45S_\gamma}\right)
\left(\frac{2\pi kT}{m_uc^2}\right)^{3/2}
\exp\left[\frac{B(Z,A)}{(A-1)kT}\right],\label{eq-nse2}\\
&\equiv&\exp\left[\frac{B(Z,A)}{(A-1)kT}-\frac{B_0}{kT}\right],
\label{eq-nse3}
\end{eqnarray}
where
\begin{equation}
S_\gamma=\frac{4\pi^2}{45}\left(\frac{kT}{\hbar c}\right)^3\left(\frac{m_u}{\rho}\right)
=1.21\frac{T_{10}^3}{\rho_8}
\label{eq-sg}
\end{equation}
is the entropy of photons in units of $k$ per baryon, and $c$ is the speed of light. 
In the above equation, $T_{10}$ is $T$ in units of $10^{10}$~K and 
$\rho_8$ is $\rho$ in units of $10^8$~g~cm$^{-3}$.
Similar notation will be used hereafter. The quantity
$B_0$ defined by Eqs.~(\ref{eq-nse2}) and (\ref{eq-nse3}) can be evaluated as
\begin{equation}
B_0=T_{10}\left[7.36+0.862\ln\left(\frac{S_\gamma}{T_{10}^{3/2}}\right)\right]\ {\rm MeV}.
\label{eq-bind}
\end{equation}
It can be seen from Eqs.~(\ref{eq-nse1}) and (\ref{eq-nse3}) that
the NSE abundances of nuclei with $B(Z,A)/(A-1)<B_0$ are exponentially suppressed.

The highest value of $B(Z,A)/(A-1)$ among all nuclei is 9.43~MeV for $^4$He 
($\alpha$ particle). The highest value of $B(Z,A)/(A-1)$ at a specific $A$ increases 
approximately
monotonically from 8.38~MeV for $^{12}$C to a peak at 8.95~MeV for $^{56}$Fe,
and then decreases approximately monotonically, reaching $\sim 8$~MeV
for $A\sim 190$ and 7.60~MeV for $^{238}$U. 
For fixed $T$, $B_0$ increases with increasing $S_\gamma$ 
[see Eq.~(\ref{eq-bind})]. As $B_0=9.43$~MeV for $T_{10}=1$ and $S_\gamma=11$,
the NSE abundances of all nuclei are exponentially suppressed and
the NSE composition predominantly consists of free nucleons for 
$T_{10}\sim 1$ and $S_\gamma\gtrsim 10$.
Physically, this may be explained as follows. The number of photons per baryon can be
calculated from the black-body spectrum to be $S_\gamma/3.6$. 
For $S_\gamma\gtrsim 10$, the high-energy tail of the spectrum for $T_{10}\sim 1$ 
contains enough photons that are capable of breaking up nuclei into free nucleons.
For fixed $T$, such photons become more scarce as $S_\gamma$ decreases.
Consequently, $\alpha$ particles start to be favored by NSE
for $T_{10}\sim 1$ and $S_\gamma\sim 1$, and both $\alpha$ particles and
heavier nuclei are favored for $T_{10}\sim 1$ and $S_\gamma\lesssim 0.1$.
In these cases, the dependence of the characteristic NSE composition on $Y_e$
is manifest. For $Y_e<0.5$ (i.e., an overall neutron richness), a significant
number of neutrons coexist with $\alpha$ particles and heavier nuclei but the
proton abundance is very low ($Y_p\ll Y_n$). The roles of neutrons and protons
are exchanged for $Y_e\geq 0.5$. In addition, for fixed $T$ and $S_\gamma$, 
the abundant nuclei for $Y_e<0.5$ are more neutron rich and heavier than 
those for $Y_e\geq 0.5$.

Now consider the adiabatic expansion from NSE for a mass element 
with a fixed $Y_e$. At time $t=0$, the mass element has
an initial composition specified by NSE for its initial temperature $T(0)$
and density $\rho(0)$. If we know $T(t)$ and $\rho(t)$ during its expansion, 
we can follow the evolution of its composition with a nuclear reaction network.
For adiabatic expansion, the total entropy $S(T,\rho,Y_e,\{Y_i\})$ of the mass 
element has a constant value $S_0$, where $\{Y_i\}$ indicates the complete
set of number fractions for free nucleons and nuclei. 
The function $S(T,\rho,Y_e,\{Y_i\})$ can be 
calculated from the thermal distributions of the contributing particles, 
which include photons, $e^\pm$, free nucleons, and nuclei 
[see Eq.~(\ref{eq-sg}) for the contribution $S_\gamma$ from photons].
Once $T(t)$ is specified, e.g., 
\begin{equation}
T(t)=T(0)\exp(-t/\tau_{\rm dyn}),
\label{eq-tt}
\end{equation}
where $\tau_{\rm dyn}$ is a constant dynamic timescale, the corresponding 
$\rho(t)$ can be obtained from $S(T,\rho,Y_e,\{Y_i\})=S_0$. Thus, given
a specific set of $Y_e$, $S_0$, and $T(t)$ [e.g., Eq.~(\ref{eq-tt}) with 
$T_9(0)\sim 10$ and some $\tau_{\rm dyn}$], the nucleosynthesis in a mass 
element adiabatically expanding from NSE can be calculated. Whether 
$r$-process nuclei can be produced depends on the set of conditions chosen.
Clearly, an $r$ process requires $Y_e<0.5$ (i.e., a neutron-rich mass element).

Typical nucleosynthesis in an expanding mass element
can be described qualitatively as follows.
In the beginning, nuclear reactions can still keep up with the NSE composition
as it shifts according to $T(t)$ and $\rho(t)$. At $T_9\sim 5$, the composition 
starts to deviate from NSE and its subsequent dynamic evolution is controlled 
by reactions involving photons and light nuclei, e.g., $(n,\gamma)$,
$(p,\gamma)$, $(\alpha,\gamma)$, $(n,p)$, $(\alpha,n)$, $(\alpha,p)$,
and their reverse reactions. At $T_9\sim 3$, all reactions involving charged
particles cease to be effective. If very few neutrons are left at this point,
the abundance pattern starts to freeze out and no $r$ process occurs. 
On the other hand, if there are many neutrons coexisting with 
heavy nuclei at this point, an $r$ process can occur subsequently with those 
heavy nuclei serving as the seed nuclei to capture neutrons. The final
abundance pattern produced by the $r$ process depends on the average mass 
number of the seed nuclei and the neutron-to-seed ratio [see Eq.~(\ref{eq-n2s})], 
both of which are set by the dynamic evolution between $T_9\sim 5$ and 3.

\subsubsection{Neutrino-driven winds from protoneutron stars\label{sec-wind}}
A well-studied model of nucleosynthesis in material adiabatically expanding
from NSE concerns neutrino-driven winds from protoneutron stars (PNSs)
produced in core-collapse supernovae (CCSNe). A PNS cools by emitting
$\nu_e$, $\bar\nu_e$, $\nu_\mu$, $\bar\nu_\mu$, $\nu_\tau$, and
$\bar\nu_\tau$. As these neutrinos pass through the hot material 
predominantly consisting of free nucleons immediately outside the PNS, 
a fraction of the $\nu_e$ and $\bar\nu_e$ can be absorbed through
\begin{eqnarray}
\nu_e+n\to p+e^-,\label{eq-nun}\\
\bar\nu_e+p\to n+e^+.\label{eq-nup}
\end{eqnarray}
The energy provided by the above reactions heats the material and enables 
it to escape from the gravitational potential of the PNS in a neutrino-driven wind
(e.g., \cite{1986ApJ...309..141D}). These reactions also interconvert neutrons
and protons, thereby determining the $Y_e$ in the wind 
\cite{1993PhRvL..71.1965Q}. As a mass element in the wind moves away
from the PNS, the rate of heating it receives decreases due to diminishing
neutrino fluxes. Eventually, its expansion can be characterized approximately
with a fixed set of $Y_e$, $S_0$, and $\tau_{\rm dyn}$. This ``initial'' state
typically corresponds to $T_9(0)\sim 10$ and an NSE composition that 
predominantly consists of free nucleons, with $Y_n(0)\approx 1-Y_e$ and
$Y_p(0)\approx Y_e$.
At $T_9\sim 5$, the NSE composition shifts to $\alpha$ particles, with
the number fraction for these particles being $Y_\alpha\sim Y_e/2$ for
$Y_e<0.5$ and $Y_\alpha\sim(1-Y_e)/2$ for $Y_e>0.5$. The corresponding
number fractions for free nucleons are $Y_n\sim 1-2Y_e\gg Y_p$
for $Y_e<0.5$ and $Y_n\ll Y_p\sim 2Y_e-1$ for $Y_e>0.5$.
An $r$ process requires $Y_e<0.5$ and that few
neutrons are consumed and few heavy nuclei are made during the
subsequent dynamic phase for setting the neutron-to-seed ratio
(prior to the cessation of reactions involving charged particles at $T_9\sim 3$).

Qualitatively, a low $Y_e$, a high $S_0$, and a short $\tau_{\rm dyn}$ favor 
a high neutron-to-seed ratio for the $r$ process.
The lower $Y_e$ is, the more neutrons are available. The shorter 
$\tau_{\rm dyn}$ is, the less time there is for consuming neutrons and 
making seed nuclei. 
Production of seed nuclei from neutrons and $\alpha$ particles must pass
through the bottleneck of $^9{\rm Be}+\alpha\to{^{12}{\rm C}}+n$, where the
abundance of $^9$Be is determined approximately by the statistical equilibrium
between the forward and reverse reactions below \cite{1992ApJ...395..202W}:
\begin{equation}
 \alpha+\alpha+n\rightleftharpoons{^9{\rm Be}}+\gamma.
 \label{eq-be9}
 \end{equation}
As discussed above, a high entropy per baryon favors free nucleons over nuclei 
in NSE. In a similar manner, a high $S_0$ suppresses the abundance of $^9$Be 
by the presence of many photons per baryon, enough of which have
energy exceeding 1.573~MeV to incur frequently the reverse
reaction in Eq.~(\ref{eq-be9}). Thus, a high $S_0$ has the same effect as
a short $\tau_{\rm dyn}$ in reducing the consumption of neutrons and the
production of seed nuclei before an $r$ process could occur.

At the end of the evolution of a massive star, it has 
a white-dwarf-like core surrounded by an envelope. The gravitational collapse
of the core initiates the CCSN process.
A neutrino-driven wind from the PNS was clearly 
seen in a model for the accretion-induced collapse of 
a bare white dwarf \cite{1992ApJ...391..228W}. 
A star of $8.8\,M_\odot$, which lies near the lower end of the mass range producing 
CCSNe, develops an O-Ne-Mg core \cite{1987ApJ...322..206N}.
In a CCSN model of such a star, Ref.~\cite{1988ApJ...334..909M} showed
that the neutrino reactions in Eqs.~(\ref{eq-nun}) and (\ref{eq-nup}) first
provide heating to the shock wave that drives the explosion and then continue
to operate to drive an expanding hot ``bubble'' between the PNS and the 
outgoing shock. The same phenomena were also seen in a CCSN model of 
a $20\,M_\odot$ star, which develops an Fe core \cite{1994ApJ...433..229W}.
Except for the effect of the shock, the hot bubble has similar 
characteristics to a neutrino-driven wind. Nucleosynthesis calculations showed
that nuclei heavier than Fe, including heavy $r$-process nuclei in some cases,
were produced in this type of environment (e.g., \cite{1992ApJ...395..202W,%
1994ApJ...433..229W,1992ApJ...399..656M,1994A&A...286..857T}). 
This inspired the neutrino-driven wind model of the $r$ process. 
However, other studies found that the winds typically have $Y_e\sim 0.4$--0.5,
$S_0\sim 10$--100, and $\tau_{\rm dyn}\sim 0.01$--0.1~s
(e.g., \cite{1994A&A...286..841W,1996ApJ...471..331Q,2001ApJ...562..887T}), 
for which only elements such as Sr, Y, and Zr with $A\sim 90$ are readily
produced. For the most favorable of these conditions, elements from
Pd and Ag with $A\sim 110$ up to Te immediately below $A\sim 130$
can be produced. In all cases, the production of elements heavier than Fe
occurs during the dynamic phase between $T_9\sim 5$ and 3, where many
types of reactions involving free nucleons, $\alpha$ particles, and photons
play important roles. Therefore, the neutrino-driven wind facilitates a new
kind of nucleosynthesis for producing elements heavier than Fe that differs
from the classical $r$ or $s$ process \cite{1992ApJ...395..202W}. However,
the general consensus is that typical conditions in the wind do not meet
the requirement for producing $r$-process nuclei with $A\gtrsim 130$
(e.g., \cite{1997ApJ...482..951H,1997ApJS..112..199M,1999ApJ...516..381F}).

The $Y_e$ in the wind is determined by the competition between the reactions
in Eqs.~(\ref{eq-nun}) and (\ref{eq-nup}). The ratio of the rates of these reactions is
\begin{equation}
\frac{\lambda_{\nu_e n}}{\lambda_{\bar\nu_ep}}=\frac{\phi_{\nu_e}}{\phi_{\bar\nu_e}}
\frac{\langle\sigma_{\nu_en}\rangle}{\langle\sigma_{\bar\nu_ep}\rangle}
\approx\frac{L_{\nu_e}}{L_{\bar\nu_e}}\frac{\epsilon_{\nu_e}+2\Delta}
{\epsilon_{\bar\nu_e}-2\Delta},
\label{eq-rye}
\end{equation}
where for example, $\phi_{\nu_e}$ is the $\nu_e$ flux, $\langle\sigma_{\nu_en}\rangle$
is the cross section for the reaction in Eq.~(\ref{eq-nun}) averaged over the $\nu_e$
spectrum, $L_{\nu_e}$ is the $\nu_e$ luminosity, 
$\epsilon_{\nu_e}\equiv\langle E_{\nu_e}^2\rangle/\langle E_{\nu_e}\rangle$ is
the effective $\nu_e$ energy, $\langle E_{\nu_e}^n\rangle$ denotes the $n$th moment of 
the $\nu_e$ spectrum, and $\Delta=1.293$~MeV is the neutron-proton mass difference.
The approximate equality in Eq.~(\ref{eq-rye}) takes into account that
$\phi_{\nu_e}\propto L_{\nu_e}/\langle E_{\nu_e}\rangle$,
$\phi_{\bar\nu_e}\propto L_{\bar\nu_e}/\langle E_{\bar\nu_e}\rangle$,
$\sigma_{\nu_en}\propto(E_{\nu_e}+\Delta)^2$, and
$\sigma_{\bar\nu_ep}\propto(E_{\nu_e}-\Delta)^2$, which also explain the definitions of 
$\epsilon_{\nu_e}$ and $\epsilon_{\bar\nu_e}$. 
A neutron-rich ($Y_e<0.5$) wind requires $\lambda_{\nu_e n}/\lambda_{\bar\nu_ep}<1$.
For $L_{\nu_e}\approx L_{\bar\nu_e}$, this requires 
$\epsilon_{\bar\nu_e}-\epsilon_{\nu_e}>4\Delta$ \cite{1996ApJ...471..331Q}.
Recent neutrino transport calculations \cite{2010PhRvL.104y1101H,2010A&A...517A..80F} 
give softer spectra than previous works (e.g., \cite{1994ApJ...433..229W}).
As a result, the neutrino-driven wind was found to be mostly proton rich
(e.g., \cite{2010ApJ...722..954R}). While this is detrimental to the prospect of an
$r$ process, other interesting nucleosynthesis can still take place in
a proton-rich wind, in particular the so-called $\nu p$ process \cite{2006PhRvL..96n2502F}.
This process goes through a phase where $(p,\gamma)$ and
$(\gamma,p)$ reactions are in equilibrium and progress towards heavier nuclei depends
on the $\beta^+$ decay of the ``waiting-point'' nuclei favored by this equilibrium 
[cf. $(n,\gamma)$-$(\gamma,n)$ equilibrium for a hot $r$ process]. 
A long-lived waiting-point nucleus would have terminated the nuclear flow were neutrinos
to be ignored (e.g., \cite{2005ApJ...623..325P}). 
However, a proton-rich wind is exposed to an intense $\bar\nu_e$ flux,
which produces neutrons through the reaction in Eq.~(\ref{eq-nup}). The ensuing
$(n,p)$ reaction on the stalling waiting-point nucleus has the same effect as its $\beta^+$ 
decay but occurs mush faster. Consequently, the $\nu p$ process can produce elements
significantly heavier than Fe 
\cite{2006PhRvL..96n2502F,2006ApJ...637..415F,2006ApJ...644.1028P}. 
It was shown that elements from Sr, Y, and Zr to Pd and Ag with $A\sim 90$--110
can be produced in both proton-rich and neutron-rich winds \cite{2011ApJ...731....5A}.

Because neutrino reactions are crucial to the determination of the conditions in the wind,
the exact nucleosynthesis in this environment is sensitive to the luminosities and spectra 
obtained from calculations of neutrino transport inside the PNS on the one hand, 
and to the effects of neutrino flavor oscillations outside the PNS on the other.
Recent works showed that medium effects increase the $\nu_e$
opacity but decrease the $\bar\nu_e$ opacity in the decoupling region, thereby increasing
the difference between the $\nu_e$ and $\bar\nu_e$ spectra
\cite{2012PhRvL.109y1104M,2012PhRvC..86f5803R}.
Consequently, $\epsilon_{\bar\nu_e}-\epsilon_{\nu_e}>4\Delta$ can be obtained to
sustain a neutron-rich neutrino-driven wind for a substantial period in the absence of
neutrino flavor oscillations. As $\nu_\mu$, $\bar\nu_\mu$, $\nu_\tau$, and $\bar\nu_\tau$
have harder spectra than both $\nu_e$ and $\bar\nu_e$, flavor oscillations outside the PNS
can drastically change the $Y_e$ in the wind (e.g., \cite{1993PhRvL..71.1965Q}).
For the benefit of the $r$ process, the most effective way to reduce $Y_e$ is to invoke
oscillations of $\nu_e$ into sterile neutrinos $\nu_s$ that do not have the normal 
weak interaction (e.g., \cite{1999PhRvC..59.2873M}). However, the presence of
multiple neutrino species with intense fluxes in an environment with a wide range 
of density calls for rather complicated treatment of flavor oscillations
(e.g., \cite{2010ARNPS..60..569D,2011JPhG...38c5201D,2012JCAP...01..013T,%
2012PhRvL.108f1101S,2012PhRvL.108z1104C,2013PhRvL.111f1101R}). 

Other than neutrino physics, the conditions in the wind depend on the mass and 
radius of the PNS. For a typical PNS of $\approx 1.4\,M_\odot$ in mass and 
$\approx 10$~km in radius, general relativity does not affect the wind very much.
Two neutron stars of $\approx 2\,M_\odot$ were discovered recently
\cite{2010Natur.467.1081D,2013Sci...340..448A}.
General relativistic effects are important for PNSs of such masses, which have
winds of substantially higher $S_0$ and shorter $\tau_{\rm dyn}$ 
favorable to an $r$ process (e.g., \cite{1996ApJ...471..331Q,1997ApJ...486L.111C}). 
Nevertheless, production of $r$-process nuclei with $A\gtrsim 130$ still requires
$Y_e\lesssim 0.4$ in the wind from a PNS of $\approx 2\,M_\odot$ 
for a reasonable nuclear equation of state
(e.g., \cite{2001ApJ...554..578W,2013ApJ...770L..22W}).
It is possible that winds from such massive PNSs can make
$r$-process nuclei with $A\gtrsim 130$ when additional beneficial
effects from flavor oscillations involving sterile neutrinos are taken into account.

\subsubsection{Other scenarios involving CCSN ejecta\label{sec-ejecta}} 
Expansion from NSE with a fixed set of $Y_e$, $S_0$, and $\tau_{\rm dyn}$
provides an approximate description for a wide range of ejecta from CCSNe.
Using the typical conditions in neutrino-driven winds from PNSs for comparison,
we can identify promising $r$-process scenarios involving other types of CCSN 
ejecta based on whether they have lower $Y_e$, higher $S_0$,
shorter $\tau_{\rm dyn}$, or any combination of the above.
Lower $Y_e$ can be obtained in winds from accretion disks (or tori) 
surrounding black holes that are formed in special CCSNe associated with long
gamma-ray bursts \cite{2003ApJ...586.1254P,2006ApJ...643.1057S}
or in neutron star mergers associated with short ones
\cite{2008ApJ...679L.117S,2012ApJ...746..180W}. The dense disk material
is neutron rich due to electron capture and emits $\bar\nu_e$ with a harder
spectrum than that of $\nu_e$. If winds are ejected rapidly without significant
neutrino interaction, they retain the low $Y_e$ of the disk material. Even
when they experience frequent neutrino interaction, the harder $\bar\nu_e$
spectrum may again lead to a low $Y_e$. So far, all $r$-process
calculations for winds from black-hole accretion disks are parametrized studies.
It remains to be seen if such studies can be confirmed by self-consistent
hydrodynamic models with accurate neutrino transport. 
Some CCSN mechanisms involve vigorous convections or 
magnetohydrodynamic jets, both of which can cause rapid ejection
of neutron-rich material near the PNS without significant neutrino interaction.
Models for $r$-process nucleosynthesis in such ejecta (e.g.,
\cite{2003ApJ...593..968W,2011ApJ...726L..15W,2012ApJ...750L..22W})
also await refinement along with the associated CCSN mechanisms.  

Extremely rapid expansion with $\tau_{\rm dyn}\sim 10^{-3}$~s may be
achieved for some shocked ejecta in O-Ne-Mg CCSNe, which is another
scenario for producing $r$-process nuclei with $A\gtrsim 130$
\cite{2007ApJ...667L.159N}. The rapid expansion required depends on
the steeply falling density profile outside an O-Ne-Mg core \cite{1987ApJ...322..206N}.
In the model originally proposed, the shocked ejecta was heated to
$T_9\sim 10$ and dissociated into free nucleons. The subsequent expansion
with $Y_e<0.5$, $S_0\sim 100$, and $\tau_{\rm dyn}\sim 10^{-3}$~s resulted
in production of heavy $r$-process nuclei up to U and Th \cite{2007ApJ...667L.159N}.
However, current models of O-Ne-Mg CCSNe do not find such conditions
\cite{2008ApJ...676L.127H,2008A&A...485..199J}.
In particular, the shocked ejecta was only heated to $T_9\sim 5$, which is too low
to dissociate nuclei into free nucleons, thereby eliminating the neutron source in
the original scenario.
So unless there would be large changes in the progenitor structure or explosion 
mechanism or both for O-Ne-Mg CCSNe relative to the current models,
shock-induced $r$-process nucleosynthesis could not work as originally proposed.
A slightly modified scenario might be more viable: if the shocked ejecta initially has 
a significant abundance of $^{13}$C or $^{22}$Ne, then a post-shock temperature
of $T_9\sim 5$ is sufficient to induce
$^{13}{\rm C}(\alpha,n){^{16}{\rm O}}$ or $^{22}{\rm Ne}(\alpha,n){^{25}{\rm Mg}}$,
which can provide neutrons for an $r$ process. These neutron sources are 
important for the $s$ process (e.g., \cite{2011RvMP...83..157K})
and were also considered in early models of shock-induced nucleosynthesis
at lower temperatures and
without rapid expansion (e.g., \cite{1978ApJ...222L..63T,1979A&A....74..175T}).
While it remains interesting to explore shock-induced $r$-process nucleosynthesis 
in O-Ne-Mg CCSNe, current models of these events were shown to be capable of
producing a number of interesting nuclei including $^{64}$Zn, $^{70}$Ge, 
$^{74}$Se, $^{78}$Kr, $^{84}$Sr, $^{90}$Zr, and $^{92}$Mo, which can be
used to constrain the frequency of occurrences for these events based on
considerations of Galactic chemical evolution \cite{2009ApJ...695..208W}.

\subsubsection{Neutron star mergers\label{sec-nsm}}
Neutron stars are a great source of low-$Y_e$ material. Such material can be 
ejected through tidal disruption during mergers of two neutron stars or of 
a neutron star with a black hole. Pioneering work on $r$-process nucleosynthesis 
during decompression of cold neutron-star matter was carried out
in Ref.~\cite{1977ApJ...213..225L}. More recent studies used detailed 
hydrodynamic simulations of mergers of two neutron stars and found robust 
production of $r$-process nuclei with $A\gtrsim 130$ 
(e.g., \cite{1999ApJ...525L.121F,%
2011ApJ...738L..32G,2012MNRAS.426.1940K,2013ApJ...773...78B}).
Based on these studies,
the extremely neutron-rich ejecta is heated by $\beta$ decay during its 
decompression and can also be shocked to high temperatures
during its dynamic ejection. Due to the very high initial density of the ejecta, heavy
nuclei are already present during the NSE phase of the expansion. The subsequent 
hot $r$ process undergoes fission cycling, thereby producing a stable abundance 
pattern for $A\gtrsim 130$. So far, the neutron-star-merger model of the $r$
process appears to be the only one that is sufficiently detailed and consistently
successful. However, due to the rather low frequency of occurrences for neutron star
mergers, this model may have difficulty explaining the $r$-process enrichment
observed in the very early Galaxy (see Sec.~\ref{sec-obs}).

\subsection{Neutrino-induced nucleosynthesis in He shells and a cold r process
\label{sec-he}}
A recent model for a cold $r$ process is related to neutrino-induced nucleosynthesis
in the He shells of early CCSNe. This is a modified version of the original scenario 
proposed in Ref.~\cite{1988PhRvL..61.2038E}, where neutrons are produced by
neutral-current neutrino reactions on $^4$He nuclei through
${^4{\rm He}}(\nu,\nu n){^3{\rm He}}(n,p){^3{\rm H}}$
or ${^4{\rm He}}(\nu,\nu p){^3{\rm H}}$ followed by ${^3{\rm H}}({^3{\rm H}},2n){^4{\rm He}}$.
These neutrons are captured by those seed nuclei (mostly $^{56}$Fe) acquired at the birth 
of the CCSN progenitor, but not by the predominant $^4$He nuclei.
If $n_n\gtrsim 10^{18}$~cm$^{-3}$ is obtained in a metal-poor CCSN with few seed nuclei,
an $r$ process can occur [see Eq.~(\ref{eq-nn})].
This scenario was critiqued in Ref.~\cite{1990ApJ...356..272W}, which concluded that
it could work only for a special class of metal-poor CCSNe with He shells
at very small radii to ensure sufficiently large neutrino fluxes for neutron production.
Some follow-up studies were carried out in 
Refs.~\cite{1998A&A...335..207N,2008JPhG...35a4061N}.

Recent models of metal-poor massive stars \cite{2002RvMP...74.1015W} show that 
a significant amount of $^{12}$C and $^{16}$O is produced in the inner He shell during 
the pre-supernova evolution. These nuclei are neutron sinks, which render the neutron 
density in the inner He shell too low to cause an $r$ process. To eliminate these neutron
sinks requires one to consider the outer He shell at radii of $r\sim 10^{10}$~cm.
This has two major effects on neutron production: (1) the large radii greatly reduce
the neutrino fluxes, and (2) the corresponding temperature of $T_9\sim 0.1$ is
high enough for efficiently producing $^7$Li through 
${^4{\rm He}}({^3{\rm H}},\gamma){^7{\rm Li}}$ but too low for the release of
neutrons through ${^7{\rm Li}}({^3{\rm H}},2n){^4{\rm He}}+{^4{\rm He}}$.
Consequently, the $^3$H nuclei produced by neutral-current neutrino reactions mostly 
end up in $^7$Li instead of generating neutrons through 
${^3{\rm H}}({^3{\rm H}},2n){^4{\rm He}}$ as proposed in the original scenario of 
a neutrino-induced $r$ process in the He shell.

A recent study \cite{2011PhRvL.106t1104B}
proposed a solution to the above problems by considering the
charged-current reaction ${^4{\rm He}}(\bar\nu_e,e^+n){^3{\rm H}}$ as
the neutron source and invoking $\bar\nu_e\leftrightarrow\bar\nu_{\mu,\tau}$
oscillations to enhance the neutron production rate.
The charged-current $\bar\nu_e$ reaction on $^4$He has a threshold of 21.6~MeV, 
which is significantly above the average $\bar\nu_e$ energy in the absence of flavor 
oscillations. However, the emission spectra of $\bar\nu_\mu$ and $\bar\nu_\tau$ 
are much harder, especially when the results from earlier neutrino transport 
calculations (e.g., \cite{1994ApJ...433..229W}) are used. 
For an inverted neutrino mass hierarchy,
$\bar\nu_e\leftrightarrow\bar\nu_{\mu,\tau}$ oscillations can occur before
neutrinos reach the outer He shell, thereby giving rise to a hard effective
$\bar\nu_e$ spectrum for neutron production. The neutron production rate 
per $^4$He nucleus is
\begin{equation}
\lambda_{\bar\nu_e\alpha,n}=\frac{1}{4\pi r^2}
\left[\frac{L_{\bar\nu_e}\langle\sigma_{\bar\nu_e\alpha,n}\rangle}
{\langle E_{\bar\nu_e}\rangle}\right]_{\rm eff}
\propto \frac{(L_{\bar\nu_e}T_{\bar\nu_e}^p)_{\rm eff}}{r^2},
\end{equation}
where $\langle\sigma_{\bar\nu_e\alpha,n}\rangle$ is the cross section
for the charged-current $\bar\nu_e$ reaction on $^4$He averaged over the spectrum,
the subscript ``eff'' denotes effective quantities for $\bar\nu_e$,
$T_{\bar\nu_e}$ is the temperature for a Fermi-Dirac 
spectrum with zero chemical potential, and the power index $p$ is
$\sim 5$--6 \cite{1988PhRvL..61.2038E,2007PhRvL..98s2501G}. 
Using the neutrino 
emission spectra of Ref.~\cite{1994ApJ...433..229W} and invoking
$\bar\nu_e\leftrightarrow\bar\nu_{\mu,\tau}$ oscillations, 
Ref.~\cite{2011PhRvL.106t1104B} showed that sufficiently high neutron 
density is obtained for an $r$ process to occur in the outer He shell of 
a CCSN of $11\,M_\odot$ with an initial metallicity of 
[Fe/H]~$\equiv\log({\rm Fe/H})-\log({\rm Fe/H})_\odot\sim -4.5$.
Nuclei with $A\gtrsim 130$ are produced on timescales of 
$\sim 10$--20~s after the onset of core collapse and remain intact after
the passage of the CCSN shock wave. The increase in density and temperature
for the shocked material also has some beneficial effects on further neutron
capture \cite{2011PhRvL.106t1104B,2013PhRvL.110n1101B}.

In the modified model for a neutrino-induced $r$ process in He
shells of metal-poor CCSNe, the newly-synthesized $^7$Li becomes
an important neutron sink \cite{2011PhRvL.106t1104B}. Interestingly, 
the high neutron density in the outer He shell not only sustains
a cold $r$ process producing nuclei with $A\gtrsim 130$, but also 
facilitates a mini-$r$ process producing $^9$Be from $^7$Li
\cite{2013PhRvL.110n1101B}. This is a signature of the above
$r$-process model as $^9$Be is usually considered to be produced 
exclusively by interaction of cosmic rays with the interstellar medium
(e.g., \cite{2012A&A...542A..67P}). Another signature is that the above 
mechanism for producing $^9$Be and $r$-process nuclei operates
only in CCSNe with initial metallicities of [Fe/H]~$\lesssim -3$.
As the metallicity increases, the neutron density obtained in the He
shell decreases, but important modification of the initial metal
abundances could still occur there, perhaps even producing
some nuclei hard to make otherwise. 

A crucial input to the above model of neutrino-induced nucleosynthesis in 
He shells is the neutrino emission spectra, especially those of $\bar\nu_e$,
$\bar\nu_{\mu}$, and $\bar\nu_{\tau}$. While modern neutrino transport 
calculations tend to give softer emission spectra
\cite{2010PhRvL.104y1101H,2010A&A...517A..80F},
they still produce significantly harder spectra for $\bar\nu_{\mu}$ and
$\bar\nu_{\tau}$ than
for $\bar\nu_e$. Consequently, even with these spectra, 
$\bar\nu_e\leftrightarrow\bar\nu_{\mu,\tau}$ oscillations can still 
greatly enhance the rate of the charged-current $\bar\nu_e$ reaction on 
$^4$He. The resulting neutron density in the He shells of metal-poor 
CCSNe is sufficiently high to produce $r$-process nuclei at least up to
$A\sim 80$ in addition to $^9$Be 
\cite{2011PhRvL.106t1104B,2013PhRvL.110n1101B}. It is also worth
emphasizing that there are still substantial uncertainties in neutrino 
transport calculations. As mentioned in Sec.~\ref{sec-wind}, 
recent studies \cite{2012PhRvL.109y1104M,2012PhRvC..86f5803R}
showed that inclusion of medium effects on the opacities of 
$\nu_e$ and $\bar\nu_e$ gives rise to significantly larger differences in
their emission spectra than those obtained by 
Refs.~\cite{2010PhRvL.104y1101H} and \cite{2010A&A...517A..80F}.
It remains to be seen whether there are further important changes, 
especially to the opacities of $\nu_{\mu}$, $\bar\nu_{\mu}$, 
$\nu_{\tau}$, and $\bar\nu_{\tau}$, 
that should be made in the current transport calculations.

\section{Implications and constraints from observations\label{sec-obs}}
Currently there is no direct observational evidence linking any astrophysical event 
to an $r$ process. However, the available data provide rich information on the
occurrences of this process and related nucleosynthesis for elements heavier than Fe. 
In this regard, elemental abundances in metal-poor stars of the Galactic halo are 
especially valuable. Those stars with [Fe/H]~$\lesssim -1.5$ were formed 
during the first $\sim 10^9$~yr since the beginning of nucleosynthesis in
the Galactic history. They survive until the present time because they
have low masses ($\lesssim 1\,M_\odot$) and hence, long lives ($\gtrsim 10^{10}$~yr). 
Low-mass stars with [Fe/H]~$\lesssim -2.5$ are particularly interesting because
they were formed at such early times that few astrophysical events could have occurred 
in a local region and chemical enrichment would be grossly inhomogeneous. 
As only short-lived massive stars could have contributed to the enrichment of the 
interstellar medium (ISM) at those early times,
low-mass stars sampling such enrichment would provide a fossil record of 
the nucleosynthesis from individual massive-star-associated sources.
A recent review of abundances in metal-poor stars can be
found in Ref.~\cite{2008ARA&A..46..241S}. Their implications for the $r$ process
were discussed in detail in Ref.~\cite{2007PhR...442..237Q}.

Based on the abundance patterns observed in metal-poor stars, the elements covered
can be approximately divided into three groups: (1) Na to Zn ($A\sim 23$--70),
(2) Sr to Ag ($A\sim 88$--110), and (3) Ba to U ($A\sim 135$--238). The variation
of the pattern within each group is typically much smaller than that of the pattern
covering any two groups. So for a qualitative discussion, we can focus on Fe, Sr,
and Eu, which represent each of the above three groups, respectively.

\begin{figure*}[hbt]
\begin{center}
\includegraphics*[angle=270, width=0.8\textwidth, keepaspectratio]{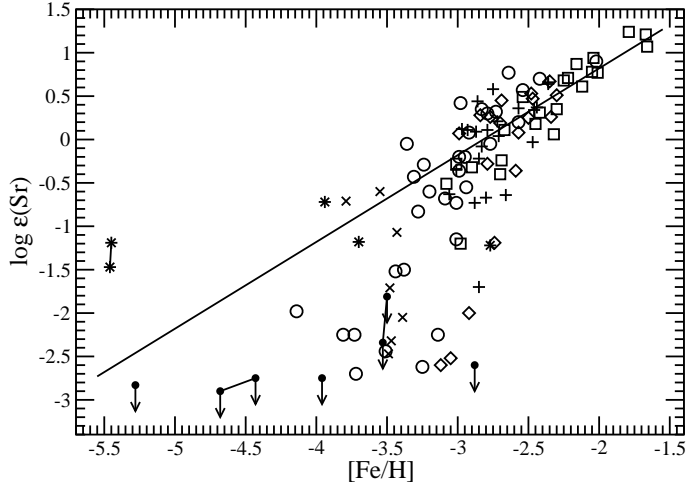}
\end{center}
\caption{Data on $\log\epsilon({\rm Sr})$ vs. [Fe/H]
(downward arrows indicating upper limits;
symbols connected with a line indicating results for the same star assuming
two different atmospheric models). Typical observational
errors in $\log\epsilon({\rm Sr})$ are $\sim 0.2$--0.3 dex.
The solid line has a slope of unity and is for an ISM with well-mixed contributions
from major sources for Sr and Fe. The data mostly cluster around this line 
at [Fe/H]~$>-2.5$ but drastically depart to low $\log\epsilon({\rm Sr})$ 
values for ${\rm [Fe/H]}\lesssim -3$. See Ref.~\cite{2008ApJ...687..272Q}
for data sources.\label{fig-srfe}} 
\end{figure*}

Figure~\ref{fig-srfe} shows the data on $\log\epsilon({\rm Sr})\equiv\log({\rm Sr/H})+12$
vs. [Fe/H] for a large sample of metal-poor stars selected in 
Ref.~\cite{2008ApJ...687..272Q}. Were there only a unique source
producing both Sr and Fe with a fixed yield ratio, the data would follow a straight line 
with a slope of unity. In sharp contrast, there is large scatter in $\log\epsilon({\rm Sr})$ 
at any fixed [Fe/H] for [Fe/H]~$\lesssim -2.5$ in Fig.~\ref{fig-srfe}, which clearly
demonstrates the occurrences of diverse sources with widely-varying relative yields
of Sr and Fe. The scatter diminishes and a clear trend emerges for [Fe/H]~$>-2.5$.
This indicates that after a sufficiently long time, contributions from all major sources
for Sr and Fe became well mixed in the ISM. In this case, both of
the number abundance ratios (Sr/H) and (Fe/H) reflect the progress of time in an
identical manner up to a constant factor. Accordingly, the data for [Fe/H]~$>-2.5$ 
follow the solid line with a slope of unity in Fig.~\ref{fig-srfe}. 
Note that relative to this line, there are many
stars severely deficient in Sr at [Fe/H]~$\lesssim -3$, which demonstrates the
occurrences of sources producing Fe but very little Sr \cite{2008ApJ...687..272Q}.
There are also quite a few stars with Sr/Fe ratios close to the solid line and one star
at [Fe/H]~$\sim -5.5$ with a very high Sr/Fe ratio. As chemical enrichment for
[Fe/H]~$\lesssim -3$ is dominated by individual nucleosynthetic events,
the relative frequencies of occurrences for sources with 
widely-varying relative yields of Sr and Fe in this regime may be inferred from
a sufficiently large sample.
Comparison of the frequencies and yields of these sources 
with the mean trend at [Fe/H]~$>-2.5$ may then inform us
if the same sources continue to be important at higher metallicities or if any of
them was only evident at [Fe/H]~$\lesssim -3$.

\begin{figure*}[hbt]
\begin{center}
\includegraphics*[angle=270, width=0.8\textwidth, keepaspectratio]{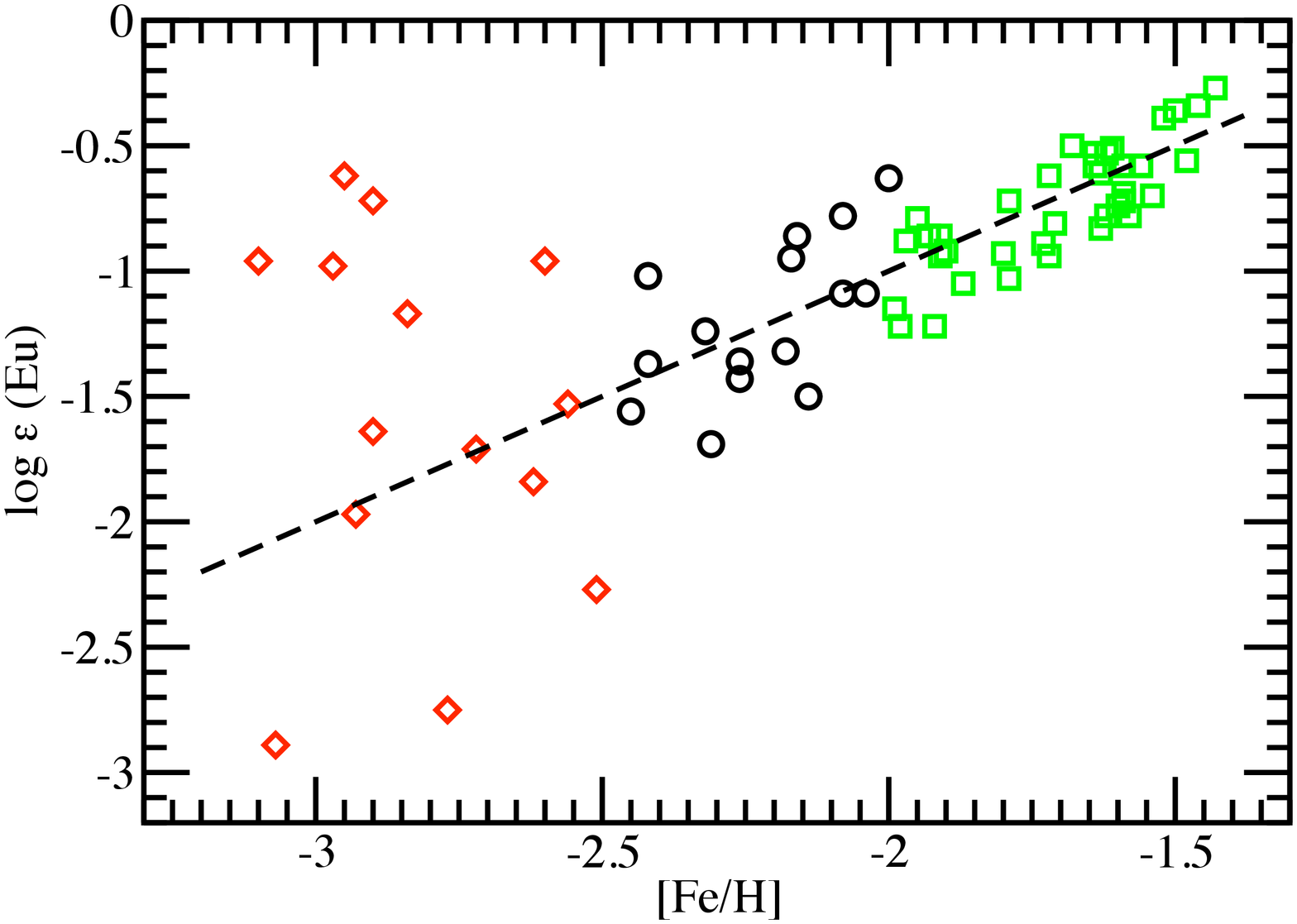}
\end{center}
\caption{Data on $\log\epsilon({\rm Eu})$ vs. [Fe/H]
(diamonds: [Fe/H]~$\leq-2.5$, circles: $-2.5<{\rm [Fe/H]}\leq -2$,
squares: [Fe/H]~$>-2$). Typical observational
errors in $\log\epsilon({\rm Eu})$ are $\sim 0.2$ dex.
The dashed line has a slope of unity and is for an ISM with well-mixed contributions
from major sources for Eu and Fe. The data mostly cluster around this line 
at [Fe/H]~$>-2.5$ but drastically depart to large scatter in 
$\log\epsilon({\rm Eu})$ for ${\rm [Fe/H]}\lesssim -3$. 
See Ref.~\cite{2010ApJ...724..975R} for data sources.\label{fig-eufe}} 
\end{figure*}

Figure~\ref{fig-eufe} shows the data on $\log\epsilon({\rm Eu})$ vs. [Fe/H]
for stars with $r$-process enrichment selected in 
Ref.~\cite{2010ApJ...724..975R}. The dashed line has a slope of unity and
approximately describes the data for [Fe/H]~$>-2.5$. However, there is
large scatter in $\log\epsilon({\rm Eu})$ both above and below this line
at any [Fe/H] for [Fe/H]~$\sim -3$ to $-2.5$. Similar to the case of Sr,
these data demonstrate the occurrences of diverse sources with 
widely-varying relative production of Eu to Fe.

Note that the lack of data on $\log\epsilon({\rm Eu})$
at [Fe/H]~$<-3$ does not necessarily mean the absence of $r$-process
enrichment at such low metallicities. This could be simply due to the
detection limit for very low Eu abundances (e.g., \cite{2013AJ....145...26R}).
As $r$-process nucleosynthesis tends to produce more Ba than Eu, Ba can
be detected down to [Fe/H]~$\sim -4$. However, there is some complication 
with Ba. Many metal-poor stars in binaries show high Ba abundances, but 
these are not due to enrichment of the ISM from which 
the stars were formed. Instead, such Ba enrichment occurred locally
in a binary long after its formation: when its more massive member evolved 
through the AGB stage, Ba was produced in that star through the
$s$ process and then transferred along with other products of AGB
nucleosynthesis to pollute the surface of the lower-mass companion, which
is now observed with a very high Ba abundance. By excluding these cases
of $s$-process contamination in binaries, we can also use Ba as
a proxy for $r$-process nucleosynthesis at low metallicities.

\begin{figure*}[hbt]
\begin{center}
\includegraphics*[angle=270, width=0.8\textwidth, keepaspectratio]{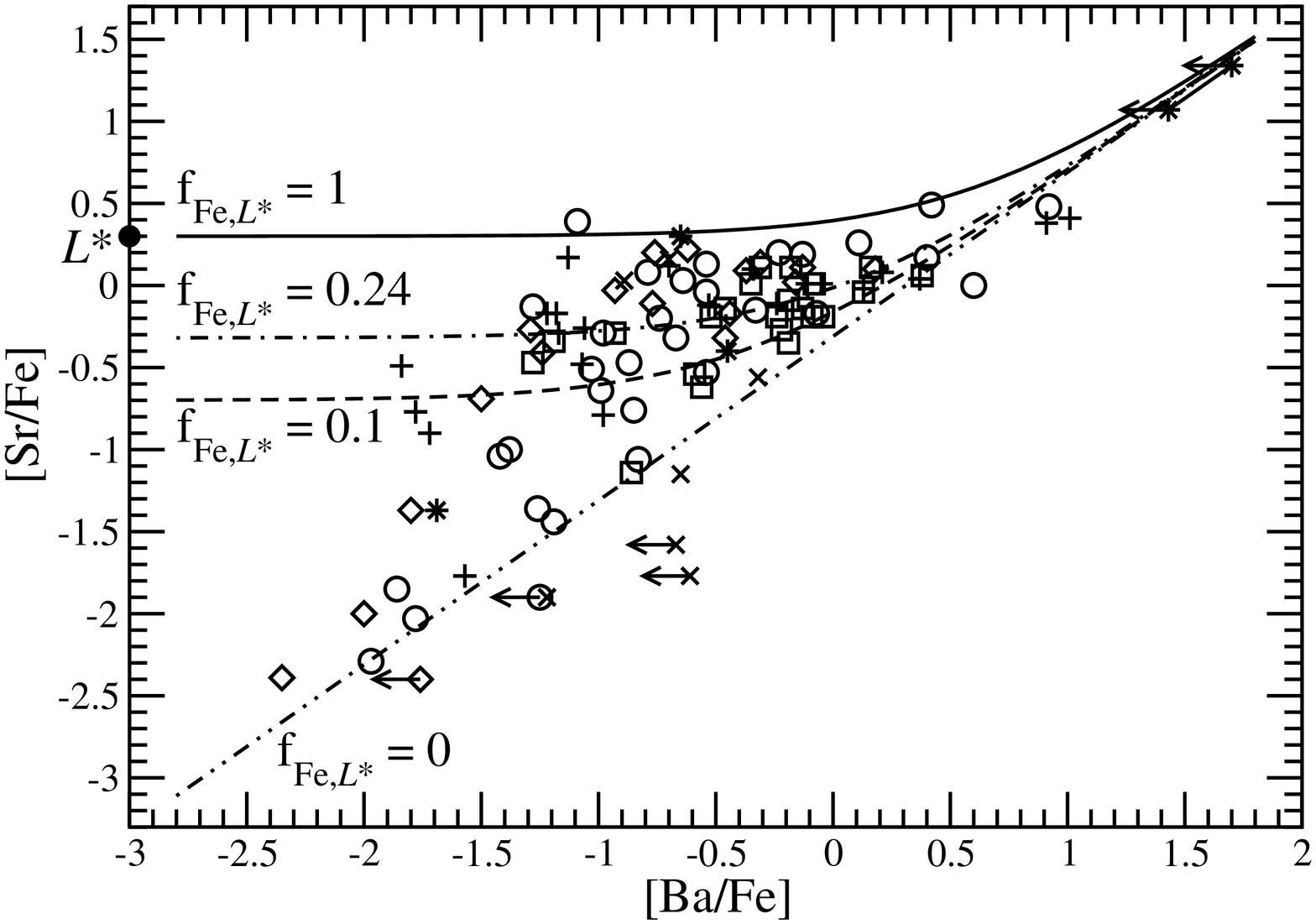}
\end{center}
\caption{Data on [Sr/Fe] vs. [Ba/Fe] for the same sample of stars
as shown in Fig.~\ref{fig-srfe} (leftward arrows indicating upper limits
on [Ba/Fe]). Typical observational errors in 
[Sr/Fe] and [Ba/Fe] are $\sim 0.1$--0.25 dex.
The curves are for an illustrative three-component model including
HNe, $H$, and $L^*$ sources.
The filled circle labeled ``$L^*$'' indicates the (number) yield ratio of
[Sr/Fe]$_{L^*}=0.30$ for the $L^*$ source.
The parameter $f_{{\rm Fe},L^*}$ is the fraction of Fe 
contributed by this source.
Note that essentially all the data lie inside the allowed region bounded
by the curves for $f_{{\rm Fe},L^*}=0$ and 1. See text and
Ref.~\cite{2008ApJ...687..272Q} for detail.\label{fig-srbafe}} 
\end{figure*}

Figure~\ref{fig-srbafe} shows the data on [Sr/Fe] vs. [Ba/Fe] for the same sample 
of stars as shown in Fig.~\ref{fig-srfe}. It is clear that diverse sources with 
widely-varying relative production of Fe, Sr, and Ba were responsible for the 
chemical enrichment of metal-poor stars. While there are a multitude of possible
ways to account for the data, it is instructive to see how a simple framework can make 
the connection between the data and the astrophysical models for nucleosynthesis.
Below we describe a three-component model \cite{2008ApJ...687..272Q} with
characteristics summarized in Table~\ref{tab-3comp} to account for 
the data in Fig.~\ref{fig-srbafe}.

\begin{table}
\caption{\label{tab-3comp}Characteristics of an illustrative three-component model.}
\begin{indented}
\item[]\begin{tabular}{@{}llll}
\br
Sources$^{\rm a}$&Stellar models&Nucleosynthetic products&Yield ratios\\
\mr
$L^*$&$\sim 15$--$20\,M_\odot$&Na to Zn, Sr to Ag& [Sr/Fe]$_{L^*}=0.30$\\
HNe&$>20\,M_\odot$&Na to Zn (especially Fe)&\\
$H$&$\sim 8$--$10\,M_\odot$, NSMs&Sr to Ag, Ba to U&[Sr/Ba]$_H=-0.31$\\
\br
\end{tabular}
\item[] $^{\rm a}$ $L^*$: CCSNe producing the group of elements associated with Fe
and the ``lighter'' part of those elements heavier than Fe including Sr to Ag. 
HNe: hypernovae
producing predominantly the group of elements associated with Fe. $H$: CCSNe or
NSMs producing both the lighter and ``heavier'' parts of those elements beyond Fe
but very little of the group associated with Fe.
\end{indented}
\end{table}

As discussed in Sec.~\ref{sec-wind}, neutrino-driven winds from PNSs are very likely 
a major source for Sr. This source may be associated with CCSNe from progenitors of 
$\sim 15$--$20\,M_\odot$, which also produce significant amounts of Fe as known from 
the light curves in some cases (e.g., Fig.~2 in Ref.~\cite{2013ARA&A..51..457N}).
It is uncertain whether the neutrino-driven winds in these CCSNe could produce
$r$-process nuclei with $A\gtrsim 130$ including Ba and Eu. In our simple framework,
we assume that they can not. For convenience, 
we will refer to the above CCSNe as the $L^*$
source producing the ``lighter'' part of those nuclei heavier than Fe and characterize
their nucleosynthesis with a fixed (number) yield ratio (Sr/Fe)$_{L^*}$.

CCSNe from progenitors of $>20\,M_\odot$ are thought to produce black holes
instead of PNSs (e.g., Fig.~2 in Ref.~\cite{2013ARA&A..51..457N}). 
At least some of these, commonly referred to as hypernovae (HNe), 
are associated with long gamma-ray bursts, which are known to produce large amounts 
of Fe (e.g., \cite{2006ARA&A..44..507W}). While winds from accretion disks associated 
with black holes formed in HNe were proposed as sites for producing a wide range of 
elements heavier than Fe (e.g., \cite{2006ApJ...643.1057S}), we assume that they 
produce very little of such elements but are a major source for Fe. 

Finally, we assume that an $H$
source is responsible for all ``heavy'' $r$-process nuclei with $A\gtrsim 130$. This source
also produces a significant amount of lighter nuclei such as Sr, but no Fe. The specific
identification of the $H$ source is uncertain, but it could be associated with low-mass
CCSNe from progenitors of $\sim 8$--$10\,M_\odot$ (e.g., \cite{2007PhR...442..237Q})
or with neutron star mergers (NSMs). Models of low-mass CCSNe showed that they 
produce very little Fe (e.g., \cite{2006A&A...450..345K}), but the neutrino-driven winds
from their PNSs can produce a significant amount of Sr. While $r$-process production
in such events is very uncertain, some modified version of a shock-induced $r$ process
cannot be ruled out (see Sec.~\ref{sec-ejecta}). It may also be possible that
a non-neutrino-driven explosion mechanism could eject very neutron-rich matter from
near the core of a low-mass CCSN to facilitate an $r$ process 
(e.g., \cite{2003ApJ...593..968W}). As for NSMs, they do not produce Fe
but were shown to make heavy $r$-process nuclei rather robustly by detailed models
(see Sec.~\ref{sec-nsm}). The neutron-rich torus surrounding black holes formed 
in these events might have neutrino-driven winds that can produce a significant 
amount of Sr (e.g., \cite{2008ApJ...679L.117S,2012ApJ...746..180W}). In any case,
we characterize the $H$ source with a fixed yield ratio (Sr/Ba)$_H$.

Taking HNe, $H$, and $L^*$ sources as operating regularly in the metal-poor regime,
we can calculate the number abundance ratio (Sr/H) in the ISM as
\begin{equation}
\left(\frac{\rm Sr}{\rm H}\right)=\left(\frac{\rm Sr}{\rm Ba}\right)_H\left(\frac{\rm Ba}{\rm H}\right)
+\left(\frac{\rm Sr}{\rm Fe}\right)_{L^*}\left(\frac{\rm Fe}{\rm H}\right)\times f_{{\rm Fe},L^*},
\end{equation}
where (Ba/H) and (Fe/H) are the abundance ratios of Ba and Fe relative to
hydrogen in the ISM, respectively, and
$f_{{\rm Fe},L^*}$ is the fraction of Fe contributed by the $L^*$ source. The above
equation can be rewritten as
\begin{equation}
10^{\rm [Sr/Fe]}=10^{{{\rm [Sr/Ba]}_H}+{\rm [Ba/Fe]}}+
f_{{\rm Fe},L^*}\times 10^{{\rm [Sr/Fe]}_{L^*}}.
\label{eq-srbafe}
\end{equation}
Using [Sr/Ba]$_H=-0.31$ and [Sr/Fe]$_{L^*}=0.30$ \cite{2008ApJ...687..272Q},
we show the curves given by Eq.~(\ref{eq-srbafe}) for $f_{{\rm Fe},L^*}=0$, 0.1, 0.24, and 1,
respectively, in Fig.~\ref{fig-srbafe}. It can be seen that essentially all the data lie within
the allowed region bounded by the curves for $f_{{\rm Fe},L^*}=0$ and 1.

The above three-component model is simple and consistent with the available data.
However, the identification of the three prototypical sources with specific models for 
nucleosynthesis is not definite. Perhaps the most uncertain identification is for 
the $H$ source responsible for all the heavy $r$-process nuclei. While current models
tend to support NSMs as a major source for such nuclei, there is a potential
problem with the data on $\log\epsilon({\rm Eu})$ vs. [Fe/H] shown in Fig.~\ref{fig-eufe}.
As NSMs occur $\sim 10^2$--$10^3$ times less frequently than CCSNe in the Galaxy
(e.g., \cite{2004ApJ...601L.179K,2004ApJ...614L.137K}), many CCSNe would have 
occurred to provide a high Fe abundance to a typical local region before it received its 
first enrichment by an NSM \cite{2000ApJ...534L..67Q,2004A&A...416..997A}.
Therefore, it would be very difficult for NSMs to account for the Eu observed at
metallicities as low as [Fe/H]~$\lesssim -2.5$ (see Fig.~\ref{fig-eufe}).
There are two possible solutions to this problem: either the frequency of occurrences
for NSMs was close to that for the dominant CCSN sources for Fe in the early
Galaxy or there were sources other than NSMs producing heavy $r$-process nuclei
at [Fe/H]~$\lesssim -2.5$. The latter solution may be provided by neutrino-induced
$r$-process nucleosynthesis in He shells of early CCSNe (see Sec.~\ref{sec-he}).
Note that apart from the above difficulty, the occurrences of NSMs over the Galactic
history appear to be consistent with the average trend of $r$-process enrichment
(e.g., \cite{2004NewA....9....1D}).

\section{Summary}
The framework of the classical $r$ and $s$ processes
proposed in Refs.~\cite{1957RvMP...29..547B} and \cite{1957PASP...69..201C} 
remains valuable in discussing the origin of the elements heavier than Fe.
However, this framework needs to be updated and complemented in many 
important aspects. As discussed in Sec.~\ref{sec-nse}, expansion from NSE 
allows a wide range of elements heavier than Fe to be produced. 
In this scenario, elements such as 
Sr to Ag can be made by reactions involving free nucleons, $\alpha$ particles,
and photons in a manner that is different from the classical $r$ or $s$ process
\cite{1992ApJ...395..202W}. Neutrino-driven winds from PNSs produced in CCSNe 
are a prototypical environment the evolution of which can be approximated by 
expansion from NSE. Here neutrinos not only determine the conditions for the 
expansion \cite{1996ApJ...471..331Q}, but also can directly
participate in nucleosynthesis as in the $\nu p$ process 
\cite{2006PhRvL..96n2502F}. Consequently, what can be produced in 
neutrino-driven winds is sensitive to the emission characteristics of neutrinos 
and to neutrino flavor oscillations (see Sec.~\ref{sec-wind}). In fact, this also 
applies to nucleosynthesis in other ejecta from 
CCSNe and NSMs so long as neutrino interaction with the ejecta is significant.
In particular, neutrino-induced $r$-process nucleosynthesis in He shells of 
metal-poor CCSNe could work only when a hard effective $\bar\nu_e$ spectrum
can be obtained before neutrinos reach the He shell (\cite{2011PhRvL.106t1104B}, 
see Sec.~\ref{sec-he}).

Nucleosynthesis studies are often challenged by the lack of accurate nuclear 
data on the one hand, and by the poor knowledge of astrophysical conditions 
on the other. This is illustrated very well by works on the $r$ process. There were 
numerous efforts at using estimates of properties of neutron-rich nuclei far from 
stability to model this process parametrically (see e.g., 
\cite{1991PhR...208..267C,2003PrPNP..50..153Q,2007PhR...450...97A}
for reviews). Without detailed knowledge of the astrophysical conditions,
many parametric studies (e.g., \cite{1993ApJ...403..216K})
calculated nucleosynthesis at a constant $T$ and 
a constant $n_n$ for a neutron irradiation time $t_{\rm irr}$, and compared 
superpositions of results obtained for different sets of $T$, $n_n$, and $t_{\rm irr}$
with the solar $r$-process abundance pattern derived from meteoritic
data and $s$-process studies (e.g., \cite{2011RvMP...83..157K}).
These calaulations provided a good guide to our understanding of the nuclear physics 
associated with the $r$ process 
(e.g., \cite{1993ApJ...403..216K,2013PhRvC..87a5805X}). 
Their adopted sets of $T$, $n_n$, and $t_{\rm irr}$
may be compatible with some astrophysical sites, but are not the result of 
self-consistent stellar models. As discussed in Sec.~\ref{sec-model}, 
important progress has been made in simulating the conditions at various potential 
$r$-process sites. Although all of these sites remain to be understood better, 
the associated nucleosynthesis studies have already benefited from the existing
qualitative, and in some cases quantitative, description of the dynamic evolution 
of the astrophysical conditions. Further, such studies can be combined with other 
characteristics of the astrophysical sites, such as their frequencies of
occurrences, to provide valuable input for a detailed framework to understand 
Galactic chemical evolution.

We have discussed a number of massive-star-associated models for
production of elements heavier than Fe in Sec.~\ref{sec-model}. Based on
the available results, it is very likely that neutrino-driven winds from PNSs
produced in CCSNe are a major source for Sr to Ag ($A\sim 88$--110). 
The associated stars are most likely of $\sim 8$--$20\,M_\odot$. With special
conditions such as a massive ($\approx 2\,M_\odot$) PNS or favorable
neutrino spectra from new physics regarding either interaction
inside the PNS or flavor oscillations outside it, the neutrino-driven winds
may be able to provide adequate conditions for an $r$ process to produce 
nuclei up to and perhaps even beyond $A\sim 130$ (see Sec.~\ref{sec-wind}).

Those elements possibly produced in winds
from PNSs may also be synthesized in other environments where expansion
from NSE occurs. Examples are winds from accretion disks (or tori) 
associated with black holes formed in CCSNe or NSMs, and
other CCSN ejecta launched from the vicinity of the PNS, which includes
convective upflows, magnetohydrodynamic jets, and
shocked inner shells with steeply-falling density (see Sec.~\ref{sec-ejecta}).
Some of the above examples, such as winds from black-hole accretion
disks, suffer from similar uncertainties in neutrino physics to the case of
winds from PNSs. In addition, all of the above examples
are subject to uncertainties in the associated progenitors
and in the dynamics involved. Nevertheless, major progress has been
made through the efforts of many workers, especially in the areas of
CCSN and NSM simulations. Stars of $\sim 8$--$10\,M_\odot$ have
a sharply-falling density profile outside the core, which makes the
associated CCSNe explode easily (e.g., \cite{2006A&A...450..345K}).
Convection plays an important, perhaps even decisive role in CCSNe 
from progenitors of $>10\,M_\odot$ (e.g., \cite{2012ARNPS..62..407J}).
Stars of $>20\,M_\odot$ are likely associated with black-hole producing
CCSNe (but see e.g., \cite{2012ARNPS..62..407J}), a fraction of these 
can be HNe to power long gamma-ray bursts (e.g., \cite{2006ARA&A..44..507W}).
The detailed nucleosynthesis associated with all these sources
remains an exciting and active area of investigation. Their neutrino signals
are also interesting to explore and may provide unique insights into the 
processes occurring deep inside these sources 
(e.g., \cite{2011PhRvD..84f3002Y,2012JCAP...07..012L}).

So far the most successful model of the $r$ process, especially regarding
production of nuclei with $A\gtrsim 130$, is associated with 
decompression of cold neutron-star matter that is ejected from NSMs 
through tidal disruption. Robust $r$-process production involving fission
cycling was seen in detailed hydrodynamic simulations of NSMs
(e.g., \cite{2012MNRAS.426.1940K,2013ApJ...773...78B}).
The data on $r$-process enrichment at [Fe/H]~$\lesssim-2.5$ appear
to require an additional source for nuclei with $A\gtrsim 130$ at such 
low metallicities.
A neutrino-induced $r$ process in He shells of early CCSNe could
fulfill this role \cite{2011PhRvL.106t1104B}. 

In summary, observations of abundances in metal-poor stars clearly
demonstrate the occurrences of 
diverse sources for elements heavier than Fe (see Sec.~\ref{sec-obs}).
The massive-star-associated models for production of such elements
discussed here not only provide a good framework to account for these
observations, but also offer many venues to explore the rich physics 
involved in stellar evolution, CCSNe, and NSMs. Most intriguing, 
the important roles of neutrinos in many of these models highlight a
profound connection between properties of elementary particles and 
the origin of the elements.

\ack
This work was supported in part by the U.S. DOE under DE-FG02-87ER40328 
and by the National Natural Science Foundation of China under Joint Research 
Grant No. 11128510. I thank Projjwal Banerjee, Wick Haxton, Alexander Heger, 
Zhu Li, Jie Meng, Zhong Ming Niu, Baohua Sun, Jerry Wasserburg, and 
Xiao Dong Xu for collaboration, 
Friedel Thielemann for helpful discussion, and two anonymous reviewers and 
Jerry Wasserburg for constructive comments and suggestions that improve
the presentation.

\section*{References}
\bibliographystyle{iopart-num}
\bibliography{ms}

\end{document}